\documentclass[useAMS,usenatbib,usegraphicx]{mn2e}

\title {Another look at Black Hole Spins} 
\author {Ari Laor, Physics Department, Technion, Israel}
\begin{document}
\maketitle

\label{firstpage}

\begin{abstract}

{\bf Do black holes rotate, and if yes, how fast? This question is fundamental and has broad implications, but still remains open.
There are significant observational challenges in current spin determinations, and future facilities offer prospects for precision measurements.}
\end{abstract}

The spin of a black hole (BH) is interesting not only ''because it is there‫''‬ (G. Mallory, 
on why climb Mt. Everest), 
but also because it tells us
about the history of the angular momentum distribution of the material from which 
the BH formed and grew, 
and the efficiency at which this material was converted to energy, and thus the potential
impact on the surrounding medium. Recent developments and future prospects on observing BH spin are comprehensively reviewed 
by Reynolds$^1$. The purpose of this comment is to mention some
pitfalls in the current techniques, and point out 
ways to get higher precision. The comment concerns mostly massive BHs in 
Active Galactic Nuclei (AGNs), where $> 99.99$\% of currently known BHs reside.

How can one measure the BH spin? The BH spin sets the radius of the innermost 
marginally stable orbit, $r_{\rm in}$, inwards of which matter cannot remain on a 
stable circular orbit, and falls freely to the BH. For a non spinning BH 
$r_{\rm in}=6r_g$ ($r_g=GM/c^2$), and for a maximally spinning BH 
$r_{\rm in}=1.23r_g$ [2]. So, measuring $r_{\rm in}$ will tell us the spin. 

How do we measure $r_{\rm in}$? Material spinning near the BH may produces 
both continuum and line emission. The Fe K$\alpha$ 
fluorescence line at 6.4~keV, inevitably produced when X-rays illuminates 
gas at $T<10^7$~K, is potentially a
particularly powerful diagnostic. The line profile is a sensitive measure of 
both the gas kinematics and the gravitational potential. Figure 1 [3] 
presents the disk line profile from the inner disk region at
$r<6r_g$, required to measure the value of the spin. For a nearly face on view
($\mu=0.99$, where $\mu=\cos \theta$, and $\theta=0$ for a face on view), 
this emission produces just a low energy tail for the K$\alpha$ line at $\sim 3.5-5$~keV
with a peak flux of only $\sim 15$\% of the peak of the outer disk line flux. 
At an intermediate inclination
($\mu=0.5$), gravitational redshift and Doppler boosting conspire to produce emission
line profiles from the inner and outer disk ($r>6r_g$) which are nearly identical. 
Only for a
close to edge-on view ($\mu=0.1$, $\theta=84\deg$) the difference becomes dramatic. 
The inner disk line is highly Doppler boosted, peaks at 9 keV, 
and drops sharply to zero at 9.5 keV. The catch is that AGNs observed at 
$\mu<0.5$ are likely
obscured by the famous obscuring torus. Such highly boosted 
lines have indeed never been observed$^4$. In most objects with disk line detections,
the profile fits give $\mu>0.5$ [4], as expected from the AGN obscuration scheme.

On top of this theoretically implied detection challenge, 
there are a host of observational challenges.
The expected line equivalent width of $\sim 150 eV$ [5], together with a typical width of 
$\sim 1 keV$ [3], implies that the line flux rises on average by only $\sim 15$\% above the continuum,
and the inner disk flux will rise by only $\sim 2-3$\% for a close to face on view. 
One therefore needs a rather
 high S/N spectrum, which is not commonly available in AGN X-ray spectroscopy.
In addition, the exact shape of the underlying x-ray continuum is not really known.
The power-law description is just a phenomenological local approximation, which breaks
below 1-2 keV, and above 10-20 keV [6], and a local spectral curvature can be mistaken as a broad
line feature. In addition, the X-ray spectral region is often rich in absorption 
and emission features$^6$, which may also make it hard to constrain the net line emission. 
Line broadening may also occur due to scattering of the photons by hot and cold electrons 
as the photons propagate out, producing blue and red line wings$^7$.

Despite all the above difficulties and systematic effects, what saves the day is
that an accretion disk line profile has a robust signature. A very sharp and steep blue
wing, and a shallow and extended red wing (Figure 1 [3]). This signature, is indeed 
clearly observed in a few high quality spectra (Figure 1 [8]), that likely prove beyond reasonable doubt that disk lines are indeed observed.  

X-ray binaries (XRBs), which host stellar mass BHs, are typically $10^4$ times brighter
in the X-rays than the brightest AGN. This provides an X-ray spectrum with effectively 
unlimited S/N (e.g. $\sim 10^6$ counts at 6-7~keV [9]). Can we measure the
BH spin accurately in these object? Formally, yes, but the complex spectral shape and 
the multitude of additional 
free parameters required to obtain a fit (26 free parameters$^9$),
makes one worry about the solution degeneracy. What is more worrying is that the 
broad K$\alpha$
profile looks like a very broad base at 5-9 keV, and does not really show the 
disk line profile signature, observed in AGNs.
This raises the suspicion that other line broadening mechanisms may be involved,
such as electron scattering, given the hot inner disk in XRBs ($10^7$K) 
compared to the cooler disks in AGNs ($10^5$K).  

Although disk line profiles can probe the BH spin, an accurate value requires
a close to edge on system, showing a highly blue shifted profile. Such a system remains to be
discovered. Can the disk continuum emission can be used instead as a more accurate probe of the BH spin?
 
An accretion disk extending below $6r_g$ inevitably produces significant continuum
emission at $r<6r_g$. In fact, if the BH is maximally spinning 
(i.e. $r_{\rm in}=1.23 r_g$) then
$\sim 80$\% of the radiation is produced at the inner disk. The five fold increase in the radiative
efficiency (from $\sim 6$\% to $\sim 30$\%), moving from a non spinning to a maximally spinning BH,
should have a dramatic effect on the spectral energy distribution (SED). However, this assumes
we know what radiation is exactly produced there. The standard thin accretion disk model
indeed provides such a prediction$^{10}$, which formally allows an
accurate determination of $r_{\rm in}$ from the SED, and thus the BH spin (once the BH mass and the disk inclination are known). The model works
well in the optical emitting region, and likely also in the near IR and near UV, but
it generally fails completely in the far UV bluewards of 1000 \AA. 
The hot inner thin disk emitting regions at $T>30,000$K, required to measure $r_{\rm in}$,
are just not there$^{11}$.  A possible physical mechanism leading
to the observed breakdown of the thin accretion disk solution, is the inevitable sharp rise 
in the gas opacity at $T>30,000$K. The opacity rise puffs up the disk, and 
may lead to a radiation pressure driven wind, as observed in O stars$^{11}$, or to
a thick configuration where the emission may be far from just a simple local 
blackbody radiation. 

Is there any way to avoid the opacity effect? If the accretion disk
surface area is large enough (i.e around more massive BHs), and the 
accretion rate is low enough, the flux per unit area may be low enough to produce  
$T<30,000$K throughout, and the UV opacity rise is avoided. Indeed, one such case
appears to be observed$^{12,13}$.  Since 
the SED solution is degenerate with the BH spin and disk inclination, the spin
can only be constrained to have low values. Clearly, followup of similar objects is worthwhile,
as this may be a particularly simple and potentially robust way to measure the spin,
in particular for the most massive BHs.

In stellar mass BH systems the accretion disk is predicted to be $\sim 100$ times
hotter, with peak emission at a couple of keV. Indeed, the SED of X-ray binaries in the soft-high-state 
is well fit by simple thermal thin disk models (up to electron scattering effects), 
which allow to constrain the BH spin$^{14}$. The surface
temperature is hot enough to eliminate almost all of the bound electrons, which
reduces the opacity to the minimal possible value of electron scattering only. 
The accretion disk likely remains thin all the way to $r_{\rm in}$. For that reason, the
classical $\alpha$-disk model was initially constructed for such systems$^{10}$,
rather than for massive BHs (R. Sunyaev, private communication). 

The BH spin in X-ray binaries does not teach us about the spins of massive BHs,
as these are formed by entirely different mechanisms (instantaneous core collapse, versus accretion over 10-100 million years). An accurate spin value can be derived
if the other system parameters, in particular the absolute luminosity, the
BH mass and the disk inclination can be constrained well by other means. 
This is a challenge for X-ray binaries, but is likely much simpler in AGNs, using
the Hubble expansion law, the broad line region, and potentially 
the continuum polarisation. 

Any other means to measure the spin?  
At $r>6r_g$ the metric is effectively independent of the spin$^2$, 
so whatever process is probed, it has to take place at $r<6r_g$
(or $r<9r_g$ for a negative spin).
Numerous various other probes were suggested, but the associated uncertainties appear
too high to derive well constrained spin values.

There are two important exceptions. The nearest massive BH at the
Galactic centre, where GRAVITY$^{15}$, and potentially also the Event Horizon Telescoe$^{16}$,
may be able to image this region directly, and eventually measure the BH spin. If these extremely powerful new technologies
will be further improved, they may allow precision spin measurements of some of
the nearest massive BHs.

The other important exception is the LIGO/Virgo experiment, 
which is already providing spin result in merging
stellar mass binary BH systems. Extension to space with LISA will allow to measure
the spins of merging massive BHs. The only caveat to this on-going revolution 
is that these results may not apply for isolated massive BHs. 

It is interesting to note that the words, {\em black hole} and {\em spin}, appear together in the abstracts of 3200 refereed papers on ADS, of which 240 appeared in 2018. This rate
of about one new paper (mostly theoretical) per work day, 
stresses the relevance of BH spins to a wide range of subjects, and the 
importance of measuring the BH spins as precisely as possible. To do that we first need to understand well the innermost 
structure of AGNs. The fruits we pick on the way on our journey towards the BH spin, 
may eventually turn out to be the real feast. \\ \\ \\
References \\ 
1. Reynolds C.~S. {\it Nat. Astron.} {\bf 3}, 41-47 (2019). \\
2. Cunningham C.~T. {\it Astrophys. J.} {\bf 202}, 788-802 (1975).\\ 
3. Laor, A. {\it Astrophys. J.} {\bf 376}, 90-94 (1991). \\
4. Nandra K., O'Neill, P. M., George, I. M. \& Reeves, J. N. {\it Mon. Not. R. Astron. Soc.} {\bf 382}, 194-228 (2007).\\ 
5. George I.~M. \& Fabian A.~C. {\it Mon. Not. R. Astron. Soc.} {\bf 249}, 352-367 (1991).\\
6. Turner T.~J. \& Miller L. {\it Astron. Astrophys. Rv.} {\bf 17}, 47-104 (2009).\\ 
7. Steiner J.~F., et al. {\it Astrophys. J.} {\bf 836}, 119-130 (2017).\\ 
8  Fabian A.~C., et al. {\it Nat.} {\bf 459}, 540-542 (2009).\\ 
9. Miller, J. M. et al. {\it Astrophys. J. Lett.} {\bf 860}, L28-L34 (2018).\\
10. Shakura N.~I. \& Sunyaev R.~A. {\it Astron. Astrophys.} {\bf 24}, 337-355 (1973).\\
11. Laor A. \& Davis S.~W. {\it Mon. Not. R. Astron. Soc.} {\bf 438}, 3024-3038 (2014).\\
12  Czerny B., Hryniewicz, K., Nikołajuk, M. \& Sądowski, A. {\it Mon. Not. R. Astron. Soc.} {\bf 415}, 2942-2952 (2011).\\ 
13. Laor A. \& Davis S.~W. {\it Mon. Not. R. Astron. Soc.} {\bf 417}, 681-688 (2011).\\
14. Davis S.~W., Done, C. \& Blaes, O. M. {\it Astrophys. J.} {\bf 647}, 525 (2006).\\ 
15. Gravity Collaboration, et al. {\it Astron. Astrophys.} {\bf 615}, L15-L24 (2018).\\
16. Zhu Z., Johnson, M. D. \& Narayan, R. {\it Astrophys. J.} {\bf 870}, 6-19 (2019).\\

 \begin{figure*}
\includegraphics[width=18cm]{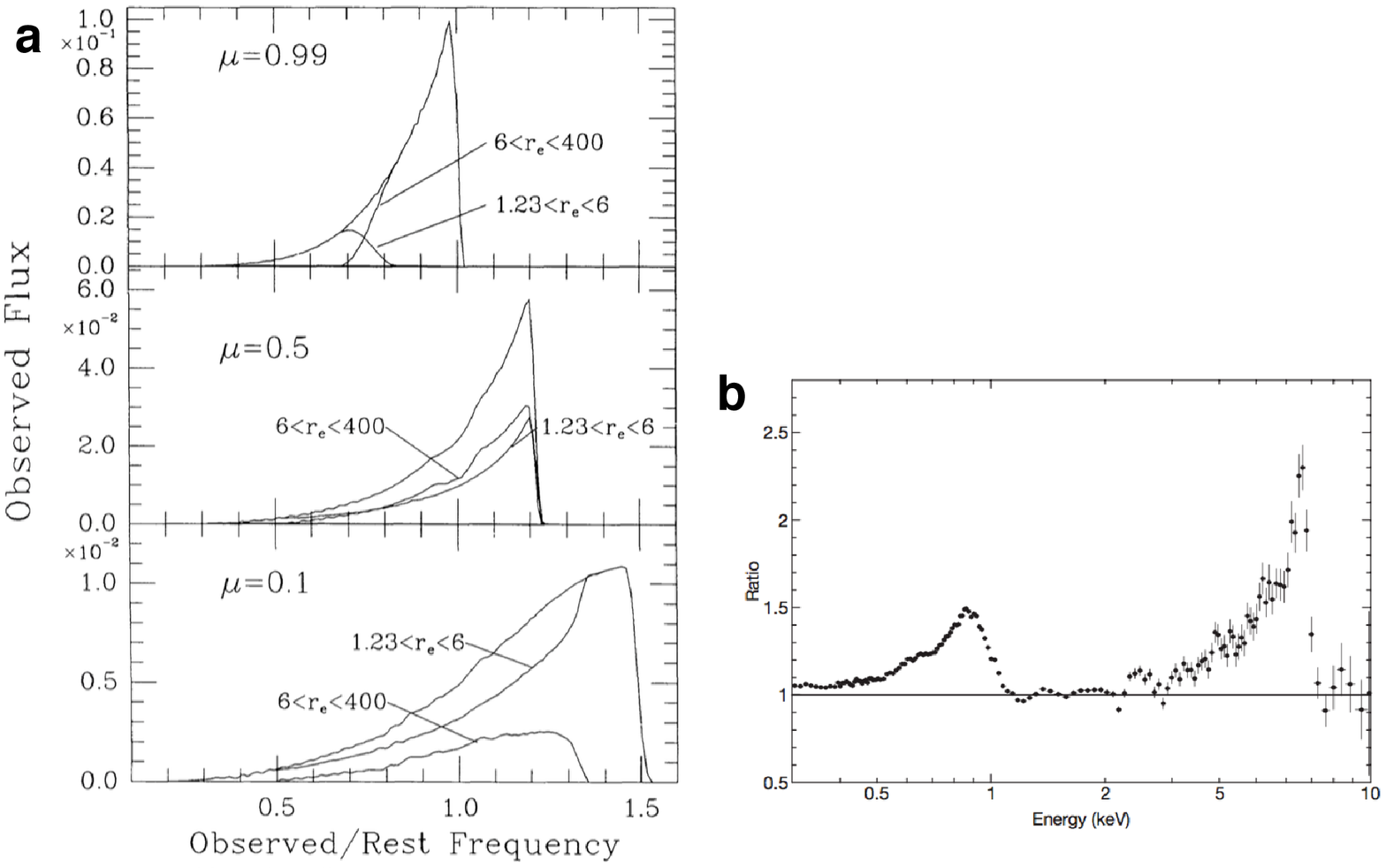}
\vspace*{-8cm}
\caption{{\bf Predicted versus observed disk line profiles. a,}
The predicted disk line profile (in arbitrary units of flux per energy bin)
from a maximally rotating BH,
observed at different inclination angles $\theta$, where $\mu=\cos \theta$,
and $\mu=0.99$ is close to face on. The contribution inwards and outwards 
of $6r_g$ are indicated, together with the total emission [3]. 
{\bf b,} A ratio plot of the observed continuum
in 1H 0707-495, divided by a standard continuum model [8]. Strong
Fe K$\alpha$ and L$\alpha$ lines are clearly detected. 
The line profiles are remarkably similar to the predicted disk line profiles,
and indicate a maximally rotating BH observed at $\mu=0.56$ [8]. Note that at this
inclination the profiles from inwards and outwards of $6r_g$ happen to be
similar (see panel a), and it may be difficult to constrain well 
$r_{\rm in}$, and thus measure the spin accurately.}

\end{figure*}

\end{document}